# Experimental realization of the ground state for the antiferromagnetic Ising model on a triangular lattice

Ke Wang[1#], Xing-Jian Liu[1#], Li-Ming Tu[1], Jia-Jie Zhang[1], Vladimir N. Gladilin[2], Jun-Yi Ge[1,3*]

1. Materials Genome Institute, Shanghai University, 200444 Shanghai, China
2. TQC - Theory of Quantum Systems and Complex Systems, Universiteit Antwerpen; B-2610 Antwerpen, Belgium
3. Department of Physics and Shanghai Key Laboratory of High Temperature Superconductors, 200444 Shanghai, China

# These authors contributed equally
*Corresponding author: junyi_ge@t.shu.edu.cn

**The antiferromagnetic Ising model on a triangular lattice (AFIT) exemplifies the most classical frustration system, arising from its triangular geometry that prevents all interactions from being simultaneously satisfied. Understanding geometric frustration in AFIT is crucial for advancing our knowledge of materials science and complex phases of matter. Here, we present a simple platform to study AFIT by arranging cylindrical magnets in vertical cavities of a triangular lattice, where magnets can slide along the cavity axis and stabilize either at the bottom or at the top of the cavity, analogous to the bistability of the Ising spin. The strong interactions of the magnets and the unique growing process allow the frustrated behavior and its ground state configurations to be directly observed. Notably, we observe a curved stripe phase, which is exotic to the Ising model. An effective thermalization process is developed to minimize the interaction energy, facilitating the evolution of various magnetic states, thereby visually realizing the ground state antiferromagnetic Ising model. Theoretical simulations and machine learning are performed concurrently to reveal the ground state and its evolution under effective thermal fluctuations, which are remarkably consistent with experimental results. Our system provides a unique platform to study frustrated systems and pave the way for future explorations in complex geometries.**

Frustration is prevalent in diverse systems ranging from neural networks[1] to disordered solids[2], and can also arise from geometric incompatibility in crystals[3]. Among all systems with geometric frustration, the most classical one is the antiferromagnetic Ising model arranged on a triangular lattice[4], where two of the three nearest-neighbor spins satisfy the antiferromagnetic interaction, while the third cannot be antiparallel to the other two simultaneously. Wannier[4] demonstrated that frustration in triangular Ising nets leads to highly degenerate ground states. Despite the intensive theoretical studies, only limited experimental approaches have been realized. A colloidal system analogous to the antiferromagnetic Ising model was reported to exhibit zigzag stripes and subextensive entropy[5]. However, in this case the lattice deforms in order to maximize the free volume of colloidal particles, making it impossible to achieve the ground state predicted for an ideal triangular lattice. This limitation prevents further insights into the ideal Ising model, leaving the nominal ground state unexplored experimentally. Other questions, such as to what extent the theoretical ideal Ising model can agree with experimental observations, remain unanswered.

Geometric frustration, though not confined to a particular scale, predominantly emerges in the microscopic and mesoscopic domains[6]. Artificial spin ice, initially developed to mimic the spin behavior of the crystalline counterparts, consists of lithographically patterned arrays of interacting magnetic nanoislands (mimic spins) designed to study the geometric frustration in a controlled manner. This approach allows unconventional states of matter to be visualized directly in real space[7-10]. Beyond the commonly studied ferromagnetic nano-islands, intriguing physical phenomena induced by frustration can also be explored using interacting colloids (mimic pseudospins) in microstructures[11-13]. This allows for the use of video microscopy to study ordered structures and their dynamics. However, nanomagnets and colloidal particles are highly affected by temperature, and the high frustration temperature prevents the system from reaching the desired ground state[5]. Additionally, the (pseudo) spins in above mentioned ice systems are all confined in-plane, which is fundamentally different from the Ising antiferromagnet.

Here, we present an approach using macromagnets confined in cavities and arranged in a perfectly rigid triangular lattice. The magnets' up and down states are analogous to the two states of Ising spins. This system overcomes the limitations of artificial spin ice and particle ice systems, allowing for more intuitive and essential observation of the ground states. An effective thermalization process is designed to track the dynamical evolution from excited high energy states to the ground state. Theoretical simulations further confirm the ground state, where the elemental motifs and their correlations are revealed through machine learning. Macro-magnetic ice lattice can be artificially designed in shape and size, permitting manipulation and tuning at the single-particle level. This allows for the study of the statics and dynamics of pre-designed frustrated lattices and provides guidance for designing novel magnetic memory devices based on frustrated spin states.

**Design of the AFIT magnet system**
The magnets with the same magnetization are placed into a triangular lattice of cylindrical cavities (**Fig.1a**, **Methods**, **Extended Data Fig. 1**). The cavity height is larger than the length of magnet, allowing it to move along the trap and stay either at top or bottom position due to the collective magnetic interactions. Each magnet-cavity unit can be used to mimic an Ising spin with up and down states indicated by arrows in **Fig. 1b**. For each pair of magnets, a high-energy metastable state is formed when both magnets locate at the same height position (**Fig. 1c**, $Z=0$). Stable states with lower energy can be achieved by shifting one of the magnets a distance $Z$ from the other. The frustration arises from the fact that it is impossible to arrange all pairs of magnets into the low energy states simultaneously in a triangular lattice (**Fig.1b**). High-energy states occur when three magnets in a triangle are simultaneously up or down. The interaction strength between the magnets is adjusted by varying the inter-magnet spacing *L*. One unique advantage of our design is the rigid triangular lattice, which does not change with the interaction strength as distinct from what happens in the colloid system[5]. Moreover, the macro-magnet system makes it possible to introduce individual interaction units to the system one by one, an approach similar to growing a crystal from individual atoms, facilitating the minimization of interaction energy both locally and globally. Thus, the ground state is expected to be readily observed.

**Growing AFIT crystalline states**

The samples with different inter-magnet distances were grown by inserting the magnets into the cavities one by one, each magnet spontaneously choosing to form the up or down state. In general, when considering only the magnetic interaction, the number of up and down magnets should be the same. However, we have found two distinct regions from the statistics of up and down magnets for all the as-grown samples (**Fig. 2a**). Above $L = 15$ mm, the magnetic interaction becomes too weak to push the magnets at the up positions. More and more magnets prefer staying at the bottom of the cavity due to the gravity. Below $L = 15$ mm, the magnetic interactions dominate over the gravity. This is confirmed by turning the whole sample upside down, where no magnet position change is noticed. In subsequent experiments, we focus on samples with *L* below 15 mm. For the three interacting magnets in one triangle, when two of them form an up-down configuration, moving the third magnet along the cavity from top to bottom (or from bottom to top) necessitates crossing an energy barrier (inset of **Fig. 2f**). As a result, the number of magnets in the up state (the side where magnets were inserted from) is slightly higher. With decreasing *L*, the effect of such barrier is weakened as compared to the dramatically increasing magnetic interaction strength, so that the up/down ratio progressively approaches 1:1 (**Fig. 2a**). This is also consistent with the proportion evolution of the excited states (three-up or three-down) for different *L* samples (**Fig. 2b**). Thus, the intrinsic low-energy state is expected to form in the smallest *L* sample. As shown in **Fig. 2c**, a large portion of ordered phase with alternating magnet states (up: bright circle; down: dark circle) is observed for *L*=7 mm sample (**Fig. 2c**). **Figures 2d-2e** show the processed images for the as-grown sample displayed in **Fig. 2c**. In a triangular lattice, each magnet is surrounded by six nearest neighbors, which can either form satisfied (antiferromagnetic, up-down) or frustrated (ferromagnetic, up-up or down-down) bonds with the central one. By connecting the frustrated bonds of neighboring magnets with red (up-up pair) and green (down-down pair) lines, one observes single-line labyrinths (**Fig. 2d**), corresponding to the arrangement with alternating magnet positions. Delaunay triangle pattern with the filled triangles representing the excited state (red: three up; green: three down) is shown in **Fig. 2e**. In samples with large *L*, domains of frustrated bonds corresponding to disorders appear (**Extended Data Figs. 2e-2h**).

To verify whether the observed state corresponds to the lowest energy arrangement, the energy for each magnetic configuration is calculated by considering the magnetic interaction within a distance of $r < 2.1L$, which includes the nearest neighbor (NN), next nearest neighbor (NNN) and second nearest neighbor interactions. In **Fig.2f**, we show the activation energy $E_A$ needed to move a magnet in **Fig. 2c** to its opposite position. When only considering the energy difference $\delta E$ between the initial and final positions, a case similar to the Ising model, 71% of magnets in the as-grown sample already stay in the low energy state. In reality, when taking into account the energy barrier for moving the magnet along the cavity, $\delta E + E_P$ or $E_P$ depending on their original positions (e.g., **Fig. 2f** inset), a positive activation energy is always required. The results indicate that the as-grown magnetic phase truly occupies the low energy ground state.

To elucidate how the magnetic interaction contributes to the observed magnetic pattern, the pairwise correlation is studied. We define the correlation $C$ of the NN and NNN magnet pairs, such that $C = -1$ if two magnets are in the opposite positions, and $C = 1$ if two magnets are in the same position. If the configuration is uncorrelated to the lattice, the average correlation value should be zero[14]. As shown in **Extended Data Fig. 3**, the NN magnet pairs exhibit the strongest correlation, playing a dominant role in the interaction. To achieve the lowest energy, the system favors configurations where $C_{NN}$ is negative, with magnets occupying opposite positions. The NNN pairs exhibit, albeit weaker, indirect interactions due to their connection through NN magnets. Notably, $C_{NNN}$ is positive because of the strong correlations among the six nearest neighbors forcing the NNN magnets to align with the central one. Interestingly, as the magnet spacing increases, the degree of correlation for both NN and NNN remains nearly unchanged. This consistency across different lattice spacings suggests that the arrangement in macroscopic magnet system is shaped by collective interactions involving all magnets, rather than simple pairwise-additive nearest-neighbor interactions in Ising model[5]. These collective interactions span multiple ranges and iteratively influence the system, ultimately leading to the observed patterns.

**Effective thermalization**

In artificial spin ice, an AC demagnetization protocol is often applied to access the low-energy ground state, where the sample is rotated in an in-plane magnetic field that periodically reverses its direction and decays in magnitude, a process known as magnetic annealing[15-23]. Such annealing process provides insight information about the collective interactions and dynamics of various magnetic configurations. We develop an effective thermalization process by vibrating the sample with a decreasing intensity to minimize the interaction energy and access a low-energy equilibrium steady state. First, by using an external strong magnetic field to push a big part of the magnets to the top positions in the cavities, a high energy metastable phase is obtained. Then, perturbations are applied by vibrating the sample in a protocol shown in **Fig. 3a**. During this process, the magnets in the sample bounce up and down in the cavity to adjust the energy due to the collective interactions, eventually reaching a stable low-energy state.

During the thermalization process, the magnet configurations are recorded and the corresponding sample interaction energy is calculated, which gradually decreases (**Fig. 3b**). At the same time, the correlation function gradually increases (**Fig. 3b**, **Extended Data Figs. 4a-4b**). **Fig. 3c** shows an example for $L$=13 mm, where 80% of magnets were initially forced to the up state. During the vibration, the magnets have sufficient energy to overcome the energy barriers, resulting in a significant reduction in the overall energy of the sample, where the up-to-down ratio gradually decreases and finally approaches 1:1. Of the 13 possible motifs that can be formed by six NN magnet pairs (**Fig. 3d**), five can be categorized as the ground state (GS) motifs[24], while the rest are excited states (**Fig. 3d**). The ground state motifs 3c, 2b and 2c have lower energy compared with the rest ones (**Fig. 3e**). Indeed, the vibration effectively decrease the number of excited state motifs and the proportion of GS motifs reaches 80% for $L$=13 mm after the thermalization process (**Fig. 3f**).

**Simulation**

Theoretical simulations are performed to further reveal the ground state configuration and its evolution (**Methods**). Following the same protocol as in experiments, the magnets are introduced into lattice one by one (**Extended Data Fig. 5**). Unlike the Ising model of disordered stripes in the ground state[24], the simulation results in parallel zigzag stripes as the ground state of the magnet system. When $T/T_0$ = 0, where $T_0$ is a characteristic temperature for macromagnet system, the zigzag stripes only consist motifs 2b and 2c (**Fig. 4a**). Although motif 3c has the lowest energy, it is impossible to tile a magnetic pattern solely with motif 3c; a combination of excited state motifs, such as motif 1, must present. Instead, a pattern can be completely filled with motifs 2b and 2c. The stripe phase is a result of energy minimization achieved through collective interactions, rather than the independent behavior of individual motifs. At finite temperature, the orientation of the stripes is distorted by the introduction of fluctuations, and stripes of different orientations appear with the emergence of excited states (**Fig. 4a**). As the temperature increases, the number of excited states gradually increase, while the percentage of motifs 2b and 2c decreases (**Fig. 4b**, **Extended Data Movie**). Correspondingly, the total interaction energy of the system increases (**Fig. 4c**). Note that the proportion of up and down magnets remains almost unchanged (**Fig. 4d**).

Considering the complex magnetic configurations and large number of degenerate states, we further constructed the correlation matrix of the motifs and applied the Least Absolute Shrinkage and Selection Operator (LASSO) regression methodology to the simulation results[25] (**Methods**). The mean squared error (MSE) reaches its lowest value when the regularization parameter ($α$) approaches zero, indicating that all configurations contribute to the total energy (**Fig. 4e**). The signs of the parameters for the 2b, 2c, and 3c motifs are negative, suggesting that these motifs contribute negatively to the total energy (**Fig. 4f**), consistent with the three lowest energy configurations observed in the calculation (**Fig. 3e**). The correlation matrix reveals significant multicollinearity among the configurations. Specifically, the 2b, 2c, and 3c motifs exhibit inverse correlations with all other configurations, which correspond to the three lowest energy motifs (**Fig. 4g**).

The simulation results are consistent with the experimental results. The ideal antiferromagnetic Ising model, highly degenerate with extensive entropy, is influenced in real materials by small disturbances such as long-range interactions[26], anisotropy[27], and lattice deformations[28-33], which relieve frustration and result in "order by disorder"[34]. The results in small magnet spacing samples having a local tendency towards zigzag stripe exemplifies this phenomenon (**Figs. 2c-2e**). This is a combined consequence of the collective interaction of the magnets, the effective fluctuations in the experimental process, and the effect of energy barrier. Noteworthy is the fact that the previously proposed colloidal system, in which the displacement of position due to the particles interaction causes lattice deformation and compression of the stripe phase[5], is very similar to that of the macroscopic magnet system, but it occurs in an unequally spaced lattice, and the interactions between the particles are relatively ambiguous. Macroscopic magnets in a rigid triangular lattice differ from the Ising model experimentally in the first instance, while the up and down states of fixed-position magnets can be visualized with the naked eye, and therefore there is no chaos caused by the inability to manipulate the position of the particles in the colloidal system.

**Outlook**

Our research demonstrates the realization of the antiferromagnetic Ising model on a rigid triangular lattice, uncovering potential exotic states of matter under ideal geometric conditions. This study reveals the spin arrangements and low-energy configurations, which are essential for comprehending novel states of matter in geometrically frustrated systems. By employing macroscopic magnet systems, our research introduces an effective experimental approach to study geometric frustration. This method enables direct and intuitive observation and manipulation of frustration phenomena on a macroscopic scale. It facilitates further investigations by varying lattice types, adjusting spacing ratios, and modifying boundary conditions. The system effectively achieves low-energy states by tuning lattice constants and thermalization process, allowing for detailed studies of both static and dynamic properties. Theoretical calculations are performed concurrently to investigate the effects of energy barrier and randomness on macroscopic magnet system, and the theoretical and experimental results are remarkably consistent.

**Methods**

**Experiments**
The cylindrical NdFeB magnets used in the experiments have the height $h = 5.89$ mm, diameter $d = 3.97$ mm, and mass $m = 0.557$ g. Cylindrical cavities are etched in acrylic sheets with the thickness of 10 mm, designed with the diameter of 4.13 mm, slightly larger than the magnets, to enable the magnets move freely up and down. The samples have a size of $20 \times 20$ cm$^2$. In order to avoid the magnets jumping out of the cavity due to strong magnetic interactions, the top and bottom sides of the sample are covered by 2 mm thick PMMA plates and fixed with non-magnetic screws. All samples' images are taken with an optical camera.

The effective thermalization process is carried out on a non-magnetic vibration table with adjustable frequency from 30 Hz to 0 in constant steps, in which the amplitude is reduced from 5 cm to 0, with each step lasting for approximately 2 minutes.

**Interaction energy calculation**

Assume that whether the magnet could move in the cavity depends on the energy change of the system. The probability for a magnet moving to its opposite position has the following expression:

$$f(\delta E, T) = \begin{cases} 1, & \delta E < 0 \\ e^{-\frac{\delta E}{kT}}, & \delta E > 0 \end{cases}$$

Where,

$$\delta E = \begin{cases} E_b, & E_{opposite} - E_{original} < 0 \\ E_b + E_{opposite} - E_{original}, & E_{opposite} - E_{original} > 0 \end{cases}$$

$$E = \sum_{<\sigma i, \sigma j> = 1} V_s + \sum_{<\sigma i, \sigma j> = -1} V_o$$

$E_b$ is the work required to overcome energy barrier to move a magnet in the triangular lattice; $\sigma_i$ is ±1 for 'up' and 'down' magnets, respectively; $E$ is the total interaction energy between all nearby magnets within a distance of $r$<2.1$L$, while the subscript 'original' and 'opposite' refer to the magnet interaction energy at the original and its opposite positions, respectively. $T$ is the effective temperature to account for the fluctuations possibly existing in experiments. $V_s$ and $V_o$ refer to the interaction energy when two magnets form the frustrated and satisfied bond, respectively.

**LASSO regression analysis**

To conduct the LASSO regression analysis, we collect data of 100 samples by linear sampling of the normalized temperature $T/T_0$ from 0 to 0.4 for a system of 29×27 magnets, lattice constant $L$=7 mm, total energy varying from 689 mJ to 736 mJ. The wide range of temperature ensure the model's generalization ability. Thirteen different motifs were chosen to form the feature space, with total energy set as target variable. The relationship between the number of motifs and total sample energy was assumed to be linear:

$$E_{tot} = \beta_0 + \beta_1 x_1 + \beta_2 x_2 \cdots + \beta_n x_n + \varepsilon$$

Here, $\beta_i$ is the coefficient corresponding to the number of motifs $x_i$, $\varepsilon$ represent the error term. To reduce the number of non-zero coefficients, we introduce a tuning parameter α into loss function to control the strength of the penalty. As α increases, more coefficients shrink to zero.

The performance of the model was evaluated using the root mean square error (RMSE)

obtained from 10-fold cross-validation. The model with the minimal RMSE was selected as the optimal model. The resulting best-fit model for the total energy is given by:

$totE = 0.274 \times (x_0) + 0.110 \times (x_1) + (0.116) \times (x_{2a}) - 0.035 \times (x_{2b}) - 0.033 \times (x_{2c}) + 0.123 \times (x_{3a}) + 0.004 \times (x_{3b}) - 0.134 \times (x_{3c}) + 0.218 \times (x_{4a}) + 0.058 \times (x_{4b}) + 0.022 * (x_{4c}) + 0.292 * (x_5) + 712.2$ mJ. RMSE=0.6 mJ

Each term corresponds to a specific motif, with its coefficient indicating the contribution to the total energy.

**Correlation coefficient**

Correlation matrix Σ is defined as:

$$\Sigma = \begin{pmatrix} \sigma_{11} & \sigma_{12} & \cdots & \sigma_{1n} \\ \sigma_{21} & \sigma_{22} & & \sigma_{2n} \\ \vdots & & \ddots & \vdots \\ \sigma_{n1} & \sigma_{n2} & \cdots & \sigma_{nn} \end{pmatrix}$$

Where $\sigma_{ij} = \frac{cov(x_i, x_j)}{\sqrt{D_{x_i}}\sqrt{D_{x_j}}} = \frac{\mathbf{E}[(x_i - \mathbf{E}[x_i])(\varepsilon_j - \mathbf{E}[x_j])]}{\sqrt{D_{x_i}}\sqrt{D_{x_j}}}$, $x_i$ is the number of motifs *i*, *D* is the sample variance, **E** represents the expected value operator.

**Data availability** The data that support the findings of this study are available from the corresponding author upon request.

**Acknowledgements** The research is supported by the National Natural Science Foundation of China (Grants 12174242 and 11804217), the National Key Research and Development Program of China (Grant 2018YFA0704300), the Flemish Research Foundation (FWO-Vlaanderen) through Grant G061820N, and the Program for Professor of Special Appointment (Eastern Scholar) at Shanghai Institutions of Higher Learning.

**Author contributions** J.-Y.G. conceived the idea and designed the experiments. K.W. performed the experiments. X.-J.L. performed the simulations. V.N.G. calculated the interaction energy. J.-Y.G., K.W., X.-J.L., L.-M.T., J.-J.Z. and V.N.G. analyzed the data. J.-Y.G. and K.W. wrote the manuscript. J.-Y.G. supervised the project.

**Competing interests** The authors declare no competing interests.

**Additional information** Supplementary Information is available for this paper.

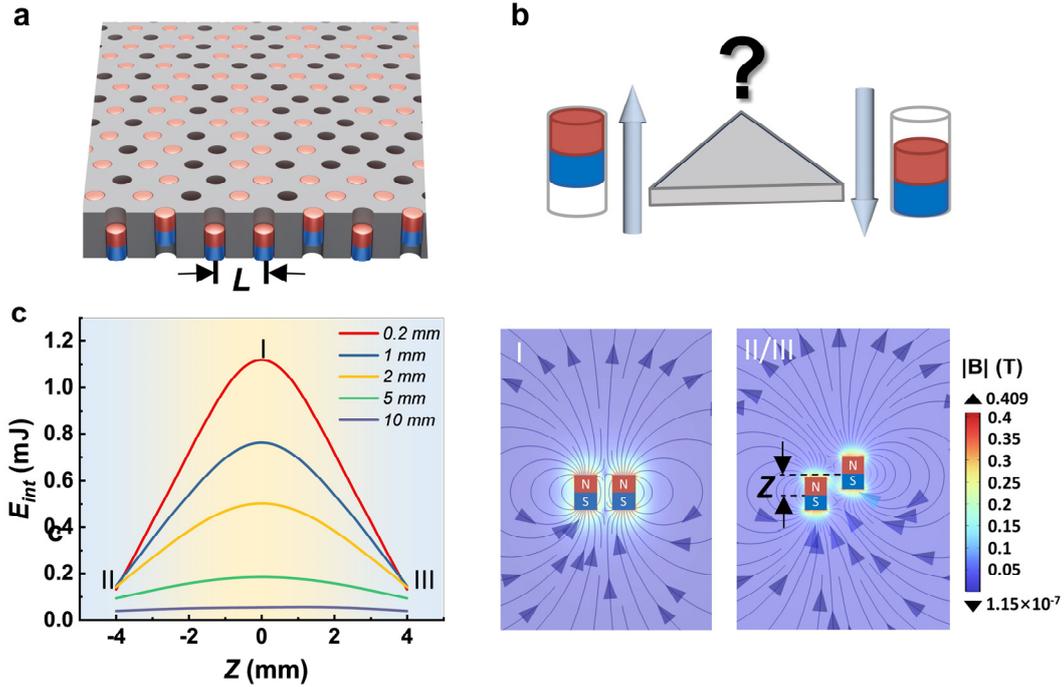

**Figure 1. The design of geometrically frustrated triangular magnet lattice**. **a,** Cylindrical magnets with the same magnetization are arranged vertically in a triangular lattice consisting of cavities. Each magnet chooses to stay either at the top or bottom of the cavity due to collective magnetic interactions. **b,** Impossibility of simultaneously satisfying all antiferromagnetic interactions in a triangle. The up and down states of magnets are analogous to the two states of an Ising spin. **c,** Interaction energy of two neighboring magnets as a function of their vertical distance $Z$ in the cavities. The up-up or down-down magnets (I, Z=0) form metastable states with high energy, a stable low energy state is achieved by shifting the magnet vertical positions to the up-down configuration (II/III, Z=4 mm). The panels I and II/III show the magnetic flux density distribution for $Z$=0 and $Z$=4 mm configurations.

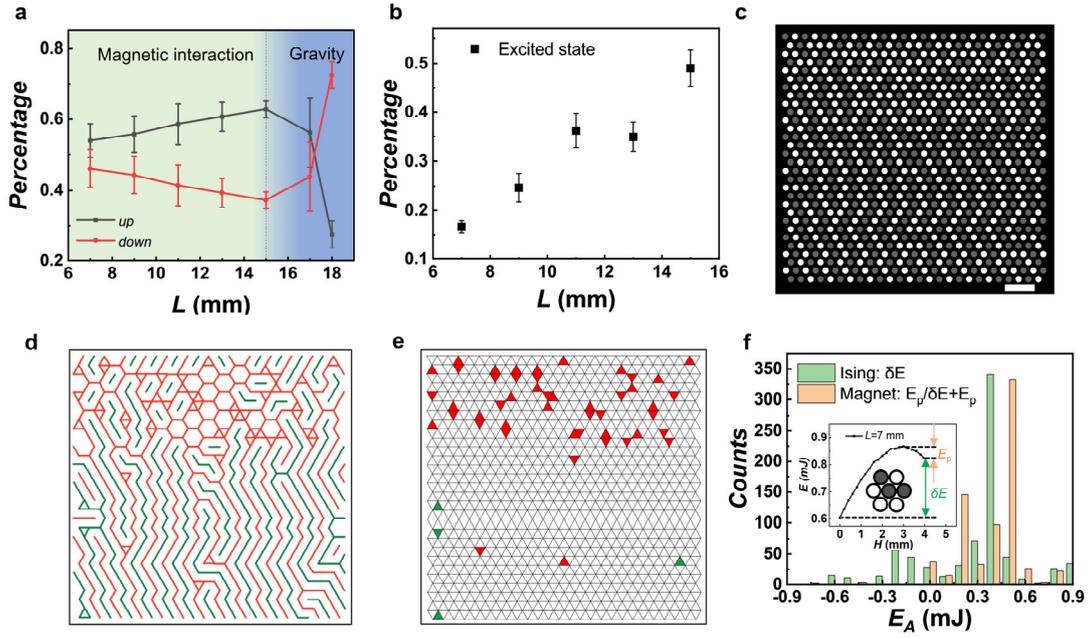

**Figure 2. Growing ground state of Ising triangular model. a,** Up-down distribution of magnets in the as-grown triangular magnet ice samples with different *L*. The regions dominated by magnetic interactions and gravity are colored green and blue, respectively. **b,** Proportion of excited states (three-up, three-down) as a function of the magnet in-plane spacing. **c,** Magnet distribution in the as-grown sample of *L*=7 mm. The up and down magnets are marked by bright and dark circles, respectively. Scale bar: 20 mm. **d,** Labyrinth patterns obtained by drawing the frustrated bonds (red: up-up; green: down-down). **e,** Corresponding Delaunay triangulation. Excited state triangles with three magnets up/down are labelled by red/green. **f,** Statistics of the activation energy needed to shift a magnet to its opposite position in **c**. The bar charts represent the two cases of whether the energy barrier, arising from the nearest neighbor interactions, is considered (green) or not (yellow). The inset illustrates an example of the energy barrier when moving the central magnet in a motif from *H*=4 mm (top) to *H*=0 (bottom) along the cavity.

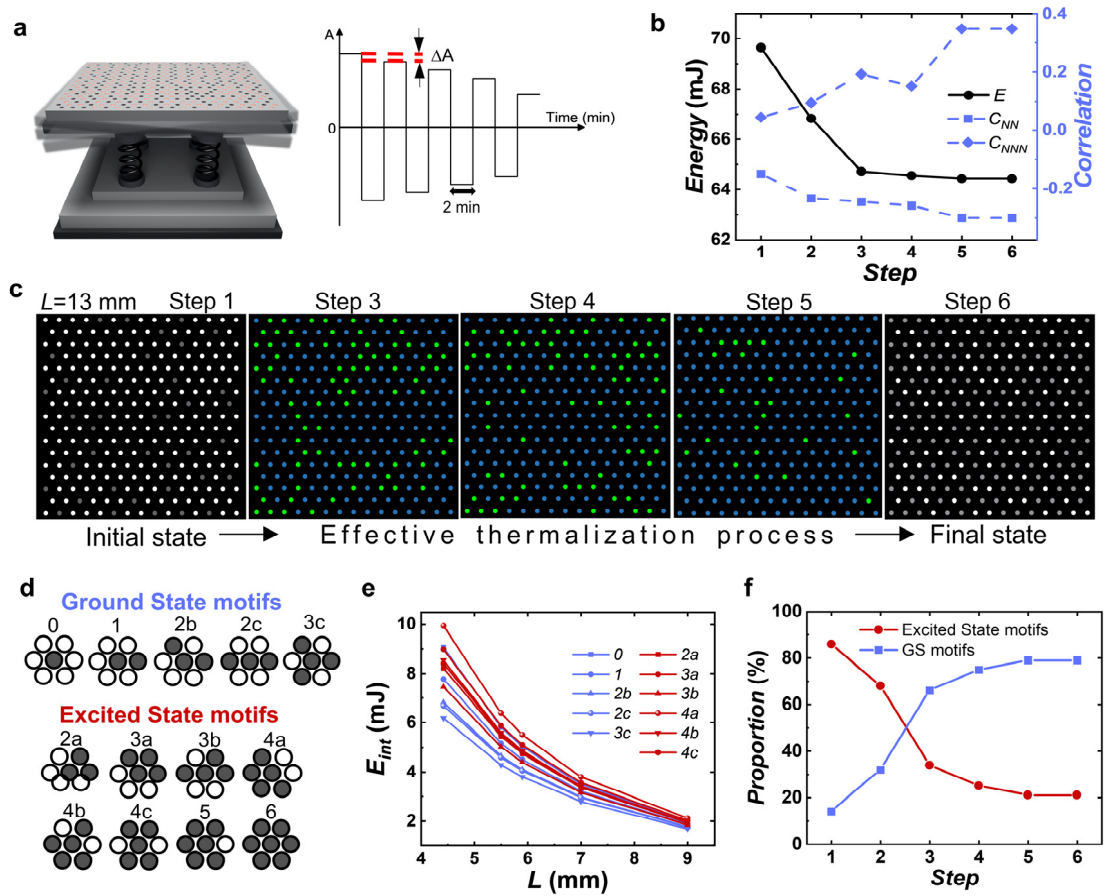

**Figure 3. Effective thermalization of high energy magnetic states. a,** Protocol used to vibrate the sample. **b,** Total interaction energy and correlation function change during the vibration process for one sample of $L$= 13 mm. **c,** Initial high energy state observed by pushing a big part of magnets to the up positions (bright circles) with a strong external magnet. After each vibration step, the magnets with changed positions are marked by green circles, while steady magnets are marked by blue circles. **d,** 13 motifs formed by six pairs of nearest neighbor magnets, which can be categorized into two categories according to the interaction energy. **e,** Interaction energy for different configurations of 7 magnets as a function of the spacing between neighboring magnets. **f,** Proportion change for the two categories of motifs during the vibration process in the $L$= 13 mm sample.

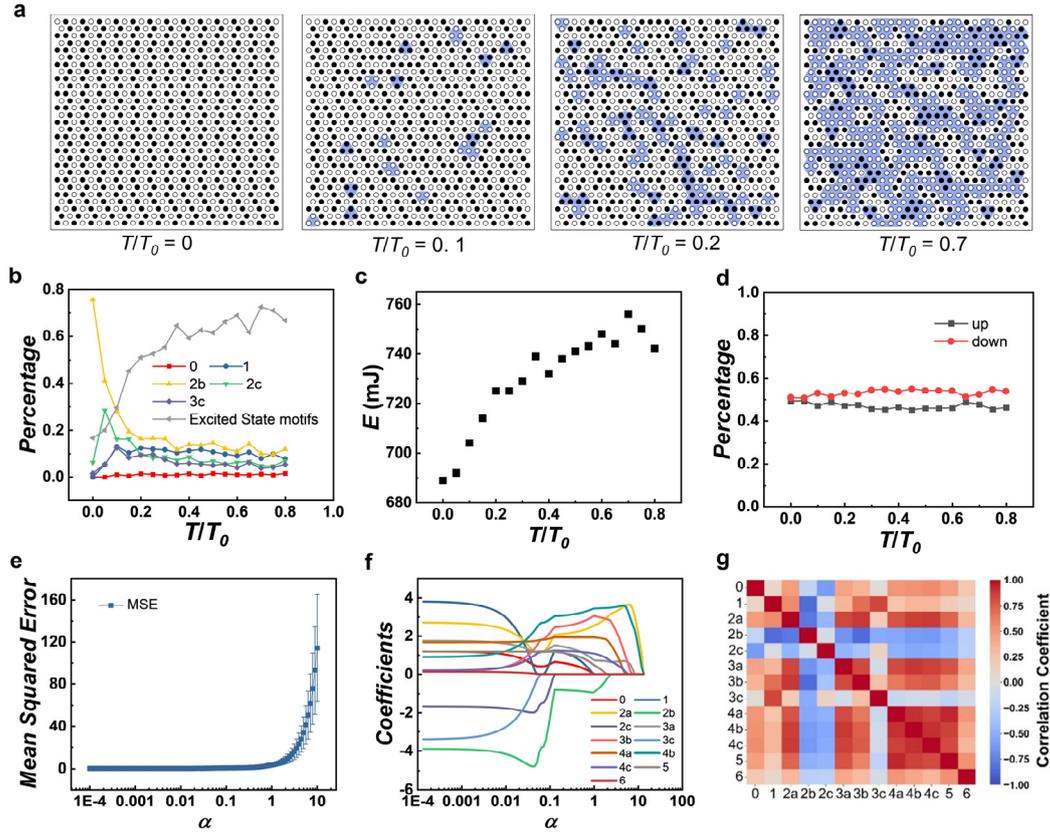

**Figure 4. Simulations. a,** Equilibrium magnet patterns observed at different effective temperatures indicated below each image for $L$= 7 mm sample. The colored areas mark the triangles with three frustrated magnet bonds (up-up, down-down). **b,** Percentage of different motif configurations as a function of effective fluctuation temperature. At $T/T_0$=0, the ground state is mainly composed with motif 2b, while motif 1 dominates at high temperatures. **c,** Evolution of total interaction energy with increasing temperature. **d,** The ratio of the number of up and down magnets remains unchanged around 1:1 with increasing temperature. **e,** The relationship between the mean square error from the result of 10-fold cross validation and the LASSO coefficient. **f,** The relationship between the configuration parameter and LASSO coefficient. **g,** The correlation matrix of different motifs that form the magnetic pattern.

# Extended Data

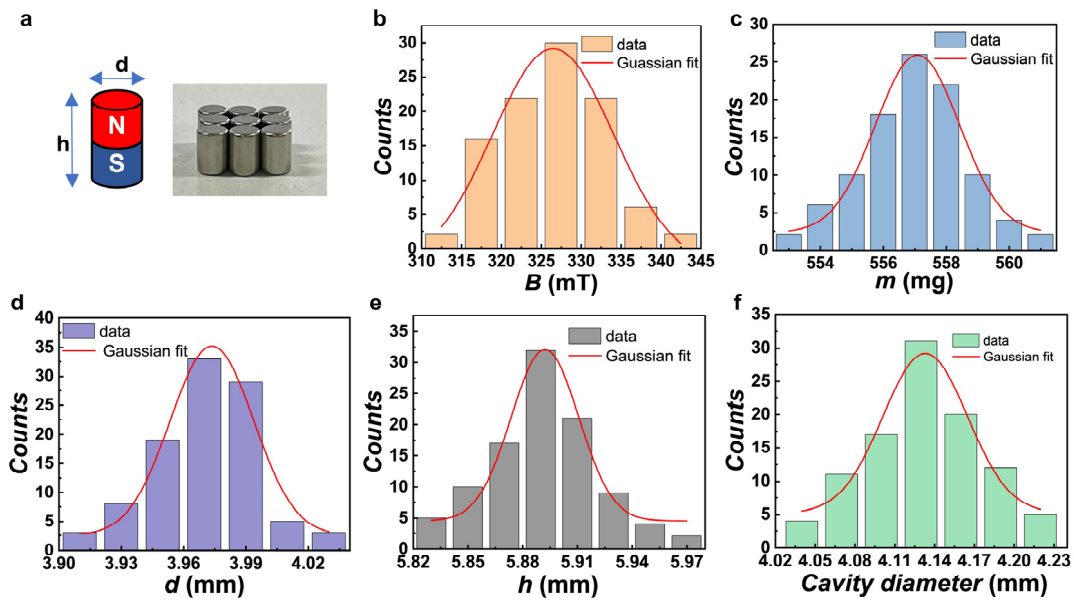

**Extended Data Fig. 1. Statistics of the parameters for magnets and cavities. a**, Schematic and optical image of the magnet. **b**, Statistics of the magnetic field measured at the magnet pole by using a Hall probe magnetic field meter, giving an average value of 326.5 mT. **c-e**, Statistics of the magnet mass (**c**), diameter (**d**) and height (**e**), yielding average values of 556.9 g, 3.97 mm and 5.89 mm, respectively. **f**, Statistics of the cavity diameter, giving an average value of 4.13 mm.

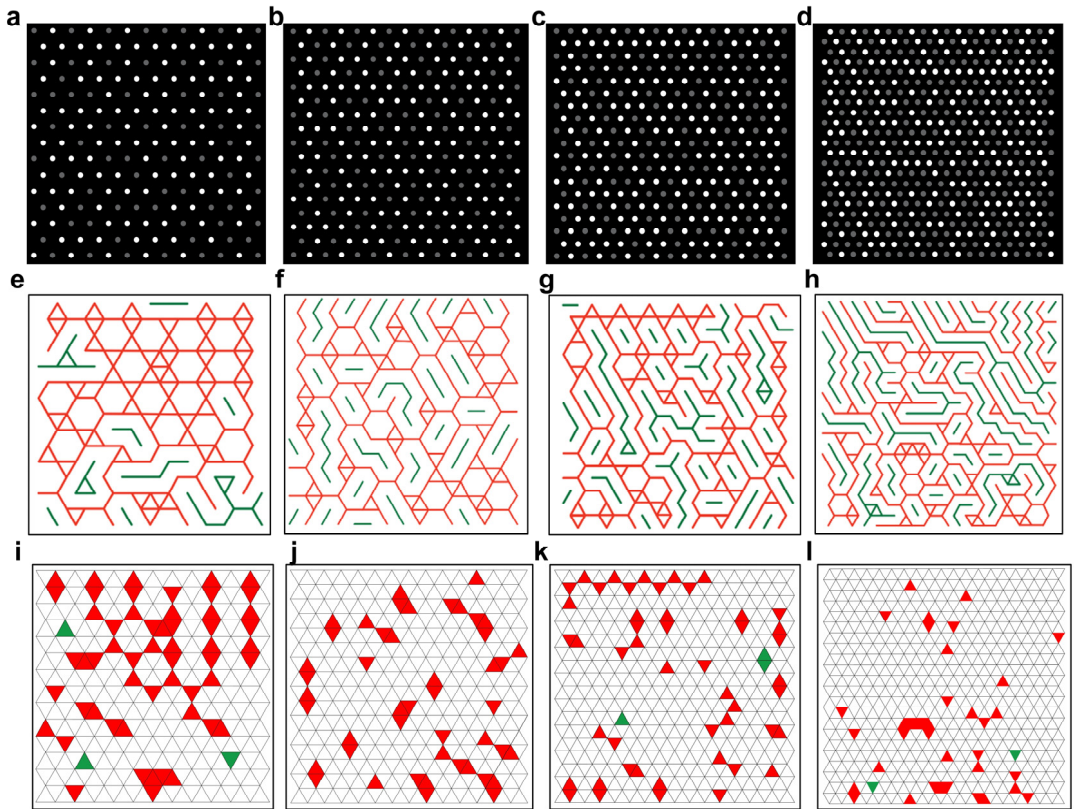

**Extended Data Fig. 2.** As-grown magnetic patterns for $L$ = 15 mm (**a**), 13 mm (**b**), 11 mm (**c**), and 9 mm (**d**). Bright and dark circles represent magnets occupying the up and down states of the cavity. **e-h,** Labyrinth patterns obtained by connecting the frustrated up-up (red) and down-down (green) bonds. **i-l,** Corresponding Delaunay triangulation. Excited state triangles with three up/down pseudospins are filled by red/green.

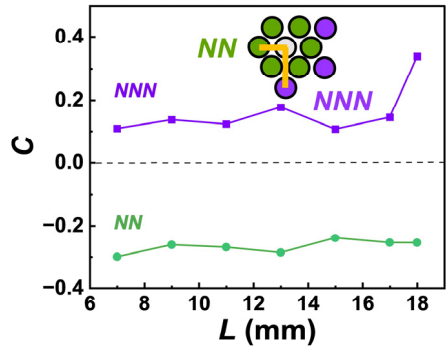

**Extended Data Fig. 3.** The correlation values between different pairs of the magnets as a function of the magnet spacing. In the inset, the nearest-neighbor magnet pairing is marked by green, while the next nearest-neighbor magnet pairing is marked by purple.

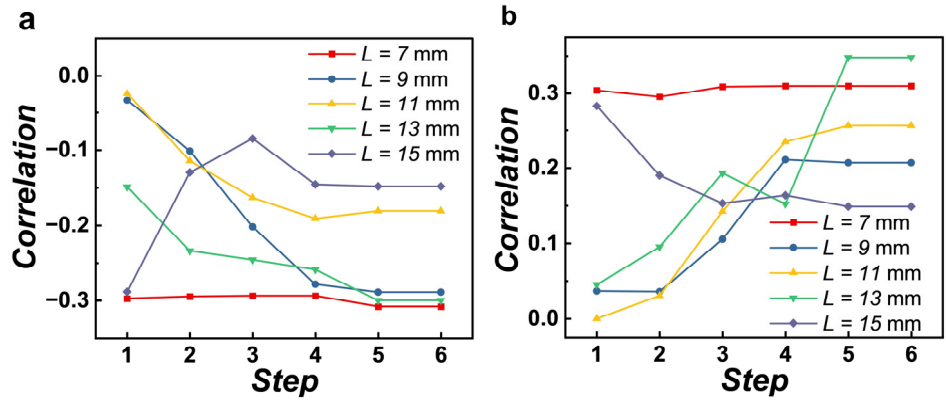

**Extended Data Fig. 4.** Changes in the correlation function for *NN* pairing (**a**) and *NNN* pairing (**b**) during the effective thermalization process. For both cases, the correlation intensity increases as the magnetic patterns becoming more and more ordered.

**Extended Data Fig. 5.** Schematic diagram showing the sequence of inserting the magnets to the lattice.

**Extended Data Movie.** Simulation of the magnetic pattern evolution with temperature increasing from $T/T_0 = 0$ to $T/T_0 = 0.7$ for $L = 7$ mm sample.